\documentclass[twoside]{article}
\usepackage{fleqn,espcrc2}

\usepackage{graphicx}
\usepackage[figuresright]{rotating}

\newcommand{\AmS}{{\protect\the\textfont2
  A\kern-.1667em\lower.5ex\hbox{M}\kern-.125emS}}
\def\be{\begin{equation}}
\def\ee{\end{equation}}
\def\bea{\begin{eqnarray}}
\def\eea{\end{eqnarray}}
\def\ii{{\rm i}}
\def\tr{{\rm Tr}}
\def\nn{\nonumber}
\def\ex{{\rm e}}
\def\lsi{\raise0.3ex\hbox{$<$\kern-0.75em\raise-1.1ex\hbox{$\sim$}}}
\def\gsi{\raise0.3ex\hbox{$>$\kern-0.75em\raise-1.1ex\hbox{$\sim$}}}
\newcommand{\lsim}{\mathop{\lsi}}
\newcommand{\gsim}{\mathop{\gsi}}
\newcommand{\lambdamsbar}{\Lambda_{\overline{\rm MS}}}
\def\bfx{{\bf x}}

\title{Static correlation lengths in QCD 
at high temperature and finite density
}

\author{Owe Philipsen\address{
                CERN Theory Division, 1211 Geneva 23, Switzerland}
        \thanks{Address since October 2000:
        Center for Theoretical Physics,
        Massachusetts Institute of Technology, 
        Cambridge, MA 02139, USA}}
       
\begin{document}

\begin{abstract}
A brief review is given of the sign problem in finite density lattice QCD and various attempts
to overcome it. To date there is still no solution to this problem which would work for
realistic QCD. The main focus then is on the deconfined phase, 
where QCD can be described by a dimensionally reduced effective
action. After summarizing derivation and validity of the effective theory, it is demonstrated
that it can be simulated efficiently in the presence of a chemical potential for quarks $\mu/T\lsim 4$.
Direct comparison of simulations with imaginary and real $\mu$ suggests that equilibrium plasma
properties could be analytically continued from 4d QCD simulations at imaginary $\mu$.
\vspace{1pc}
\end{abstract}

\maketitle

\section{INTRODUCTION}

\begin{figure}[htb]
\vspace{9pt}
\includegraphics[width=7cm]{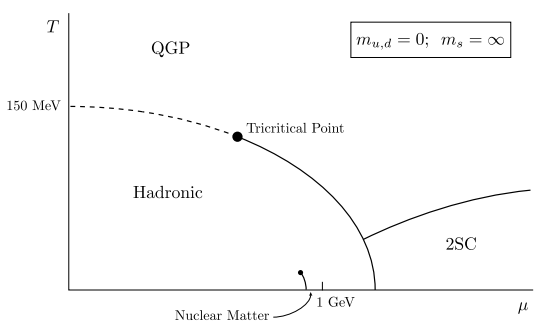}
\vspace*{-0.5cm}
\caption[]{\label{pdiag} 
Conjectured phase diagram for QCD with two massless and one heavy flavour. From \cite{kr}.}
\end{figure}

QCD at finite baryon density occurs in nature in two rather different regimes:
{\it i)} In heavy ion collisions, whose initial state has non-zero baryon number,
and in which any subsequent plasma state is a state of high temperature and low density.
{\it ii)} In the core of neutron stars, which is composed of cold and very dense matter.
These two situations correspond to the regions close to the axes in the
tentative QCD phase diagram Fig.~\ref{pdiag}.
While in the latter case some interesting conjectures about a 
colour-superconducting state of matter have been made \cite{csc}, understanding the former regime
is particularly pressing  in view of running and upcoming experiments at SPS, LHC (CERN) and
RHIC (BNL). Eventually
we would of course like to know the structure of the entire phase diagram, in particular
the position and nature of the transition lines separating different regions and 
the (tri-) critical points.

In order to come to first principles predictions, 
non-perturbative methods are required.
Unfortunately, lattice QCD has so far failed to be a viable tool for
the analysis because of the so-called ``sign-problem''. The QCD partition
function is given by
\bea
Z&=&\int DU\, \det M(\mu)\ex^{-S_g[U]},\nn\\ 
S_f&=&\int d^4x\, \sum_{colour,flavour}\bar{\psi}M\psi\nn\\
M&=&{\cal D}-\mu \gamma_0, \quad {\cal D}=\gamma\cdot(\partial
+\ii A) + m\,,
\eea
where $S_g$ is the pure gauge action.
The relation
\be \label{pos}
\gamma_5 M\gamma_5=M^\dag
\ee
implies positivity of the fermion determinant for $\mu=0$, $\det M(0) \geq 0$.
However,
for SU(3) and $\mu\neq 0$, Eq.~(\ref{pos}) does not hold, $\det M$ is complex and prohibits standard
Monte Carlo importance sampling, for which a manifestly positive measure in the functional 
integral is required.

In the first part of this contribution, 
a short review is given on the various attempts and techniques proposed in the literature to 
deal with this problem. Many of these try to implement a 
chemical potential at zero or low temperature and to obtain 
the ``onset'' value $\mu=\mu_0$, where 
the transition from
a zero density state of matter to a state of finite density takes place. At zero 
temperature this should happen when one baryon can be made, i.e.~$\mu_0=m_B/3$.
Since most of these techniques have been reviewed in detail in previous
years \cite{gla,al}, this section will be rather brief.

The larger part is then devoted to the study of the plasma situation at high temperature
and low density, where it was recently demonstrated that numerical simulations 
of static quantities
at finite density are feasible in the framework of a perturbatively
derived effective theory \cite{hlp1}. 

\section{NUMERICAL TECHNIQUES FOR \\FINITE DENSITY}
\label{meth}

\subsection{Reweighting methods}

The Glasgow method \cite{gla} evaluates the partition function by absorbing the
complex determinant into the observable, and doing the importance sampling with the
positive part of the measure,
\be
\langle{\cal O}\rangle =\left\langle \frac{{\cal O} 
\det M(\mu)}{\det M(\mu=0)}\right\rangle_{\mu=0}\;.
\ee
One may further expand the grand canonical partition function in the 
fugacity $z=\exp{\mu/T}$, 
\be
Z(\mu,T,V)=\sum_{B} z^B Z(B,T,V)\;,
\ee
where $Z_B$ is the canonical partition function at fixed baryon density, and compute
the terms individually.
However, numerical results give for the onset of baryon density the unphysical value
$\mu_0=m_{\pi/2}$ instead of the expected $\mu_0=m_{B/3}$, which is also the pathological
behaviour of the quenched theory \cite{quen}. 
It seems now understood that the reason for this failure is the poor 
overlap between the ensemble of configurations at $\mu=0$ used to compute averages
by reweighting,
and the correct ensemble at $\mu\neq 0$. 

This situation can be improved
by splitting the determinant in modulus and phase, $\det M = | \det M|\exp(\ii \phi)$,
and to incorporate the modulus into the measure, while only 
the phase is used for reweighting \cite{mod},
\be
\langle{\cal O}\rangle =\left\langle \frac{{\cal O} 
\ex^{\ii\phi}}{\ex^{\ii\phi}}\right\rangle_{|\det M|}\;.
\ee
The expectation value of the phase factor can be viewed as a ratio of two partition functions,
one of the true theory and one with the positive modulus of the determinant in its measure. 
It behaves thus as $\exp -\Delta F$, where the exponent is the free energy difference and an
extensive quantity. Consequently,
$\langle \exp\ii\phi\rangle\sim\exp(V)$,
which requires exponentially large statistics for realistic volumes.
A similar variant of the Glasgow method works for heavy quarks \cite{alo}.
 
\subsection{Static quarks}
\label{sec:stat}

The sign problem can be handled in a static quark model, in which
all quarks are infinitely heavy \cite{ben,blum}. 
It is obtained by sending both quark mass and chemical
potential to infinity $\mu,m\rightarrow\infty$, while fixing a finite 
baryon number. This is achieved by keeping the constant
\be
C=(2am/\exp(a\mu))^{N_t}
\ee
fixed, where $N_t$ is the temporal extent of the lattice.
In this case an analytic expression for the determinant is obtained which
is trivial to evaluate. The available computing time then suffices to gather
large enough statistics for the phase of the determinant to be calculated with sufficient accuracy.
The question is whether results obtained for static quarks translate at all
to QCD with small quark masses. A more detailed discussion of this approach is presented
in another contribution to this conference \cite{chan}.

\begin{figure}[htb]
\vspace{9pt}
\includegraphics[width=7cm]{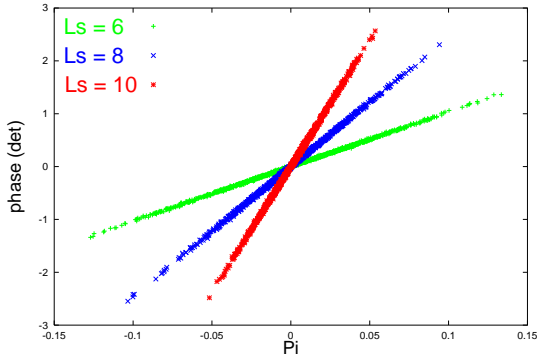}
\vspace*{-0.5cm}
\caption[]{\label{corr} Correlation between the phase of the fermion determinant and the imaginary
part ot the Polyakov line for static quarks at finite density on various volumes. From \cite{fl}.}
\end{figure}

One may attempt to extend this approach to finite but large quark masses by going beyond the 
leading term in the $1/m$-expansion, based upon an interesting observation \cite{fl}: 
analysis of the data of \cite{blum} shows a correlation between the phase of the fermion 
determinant and the imaginary part of the Polyakov line, $\phi \sim V P_i$, as shown in 
Fig.~\ref{corr}. 
Indeed, the leading order term in $1/m$ for the phase is \cite{fl}
\be \label{pp}
\phi=2(2/m)^{N_t} V \sinh(N_t\mu) P_i\;.
\ee
The correlation is weakened by higher terms in the quark mass expansion which do not
show this proportionality, and therefore 
it weakens at smaller quark masses.

This suggests the following approximation: computing a constrained
partition function 
\be
Z'=\int DU\, \det M \ex^{-S_g[U]}\delta(P_i[U])
\ee
eliminates configurations with significantly fluctuating phases $\phi$ whenever
its correlation with $P_i$ is strong, and thus 
improves the sign problem significantly. The approximation is good, i.e.~$Z'\approx Z$,
if $P_i\ll P_r$, which is the case at small $\mu/T$, or if $P_r\approx P_i\approx 0$ at 
$T\sim 0, \mu<\mu_0$. 
We shall come back to this
point when hot QCD at finite density is discussed.
 
\subsection{Imaginary chemical potential}

There are a number of suggestions to consider imaginary chemical potential, for which the 
integration measure is positive and simulations are straightforward. 
The relation to real chemical potential  
is provided by the canonical partition function at fixed quark number $Q$, which is
related to the grand canonical partition function at fixed imaginary chemical potential
by \cite{mr}
\be\label{can}
Z(T,Q) =
\int_{-\pi T}^{+\pi T} \frac{d \mu_I}{2\pi T}\, Z(T,\mu = \ii\mu_I)\,
e^{- i \mu_I Q/T}.
\ee 
One proposal is to simulate $Z$ for various values of $\mu_I$, and then numerically do
the Fourier integration \cite{aw}. This becomes more and more difficult, however, for large
$Q$, and extrapolation to the thermodynamic limit seems questionable. 
The method has been tested in the two-dimensional Hubbard model \cite{aw}, but not for QCD.

A second proposal is to compute expectation values
with the grand canonical partition function $Z(\ii\mu,T)$, 
fit them by power series in $\mu$ and analytically continue the series to real $\mu$.
This has been explored for the chiral condensate in the strong coupling limit 
\cite{lom}. 

Finally, starting from Eq.~(\ref{can}), a well-defined quenched limit of the theory at finite 
density can be obtained \cite{eng}.
The limit $m\rightarrow \infty$ is taken in the expression Eq.~(\ref{can}),
upon which it goes over into
\be
Z(T,Q) =\int DU \,f_Q[P]\ex^{-S_g[U]}\,,
\ee
where $f_Q[P]$ is a functional of Polyakov lines.
The quenched limit thus is not simply realized by replacing the fermion determinant by a constant.
Instead, there are
remainders of the static quark propagators in the partition function, the Polyakov loops, 
which carry the baryon number. 
The expression after the limit is still
complex, and thus continues to have a sign problem. However, simulations have shown
that it is weakened enough to be well manageable numerically, and results for the static potential
are available \cite{eng}.

\subsection{Avoiding the sign problem: SU(2)}

In view of the difficulties with SU(3), one may resort to studying two colour QCD 
which, thanks to the realness of the SU(2) matter representations, has a real and positive 
fermion determinant and can be simulated straightforwardly \cite{su2fund}. 
Unfortunately, the same property leads to other peculiarities which make it less analogous
to QCD than one might have hoped: 
charge conjugation is related to global SU(2) rotations. Thus, the condensates
$\langle q \bar{q}\rangle$ and $\langle qq\rangle$ are related by rotations as well, 
and there is no fundamental distinction between them. 
Since there are only two colours, `baryons' consist
of even numbers of quarks, and through SU(2) rotations 
they are equivalent to mesons. In particular, this results in
a physical onset chemical potential $\mu_0=m_\pi/2$, even in the unquenched theory. 
Moreover, since $\langle qq \rangle$ is gauge-invariant, there can be no colour 
superconductivity either. Hence, the model is simulable, but some main 
questions of interest for QCD have disappeared. 

Somewhat surprisingly, these questions reenter if the representation of the fermions is
taken to be adjoint and only one flavour is considered \cite{hands}. In this case, there
are no $qq$ Goldstone modes. Moreover, the theory permits gauge-invariant three-quark 
baryon operators $qqq$ with spin 1/2, and thus $\mu_0$ for the finite density onset 
should be different from half the meson mass. Further, $\langle qq\rangle$ is gauge-variant
and hence permits a study of colour superconductivity.
Finally, in the lattice model with an adjoint fermion, the determinant is real but not
positive definite, hence a weakend sign problem is present which 
can be handled by a two-step multi-boson algorithm \cite{hands}.

More detailed discussions of SU(2) models together with numerical 
results may be found in 
the finite density section of these proceedings \cite{su2p}.

\subsection{Finite isospin density}

In has also been proposed to consider a chemical potential
for isospin rather than for baryon number \cite{aw,ss}. 
In this case, and with two degenerate light flavours $m_u=m_d$, Eq.~(\ref{pos}) can be
supplemented by an isospin rotation to give
\be
\tau_1\gamma_5 M\gamma_5\tau_1=M^{\dag}\,,
\ee
which again implies $\det M\geq 0$ and thus can be simulated. Analytic work for very small and large
isospin chemical potential, where chiral perturbation theory or ordinary perturbation theory
are applicable, respectively, lead to the conjecture that the hadron and quark matter phases
are smoothly connected \cite{ss}. 

\subsection{Hamiltonian formulation}

Another attempt addresses the problem in a Hamiltonian
formulation of lattice QCD at finite density \cite{ham}. 
Quark number density, chiral condensate, some meson and the nucleon masses have been
calculated in the limits of a free and a strongly coupled theory at zer temperature. 
The chiral phase transition
is found to be first order, and the onset $\mu_0$ is consistent with $m_B/3$.
So far, however, no numerical implementation of this approach is available.

\subsection{Cluster algorithm}

Finally, a new algorithm has been proposed to actually solve the sign problem \cite{uwe}.
So far, it has only been worked out for non-relativistic spin systems. 
The approach uses cluster algorithms to decompose a fermion configuration into
clusters of spins that can be flipped independently, individually changing sign or 
not under the flip. 
This enables to cancel 
negative against positive contributions, leaving only net positive contributions.
Details and possible extensions
to QCD are discussed in the contribution \cite{chan}.

\subsection{Finite temperature and density}

All proposals that have been implemented numerically have thus failed
on realistic QCD so far, and the failure is particularly drastic at low temperatures.
The methods that do work numerically are limited to static or very heavy fermions,
and they work better at high temperatures, suggesting that the sign problem may
be easier to overcome in hot QCD.
Let us then shift attention to temperatures above the deconfinement transition.

\section{EFFECTIVE THEORY APPROACH TO HOT QCD}

For temperatures larger than a few times the deconfinement temperature $T_c$, 
the static (equilibrium) physics of QCD can be described by an effective theory, 
which indeed permits
simulations with non-vanishing real chemical potential \cite{hlp1}. 
Of course, the restriction to the 
deconfined phase presents a limitation. Another one is that
finite density simulations are possible only up to some value of the 
chemical potential, where the sign problem sets in. Presently the accessible range
is $\mu\lsim 4 T$. Finally, if the effective theory is to be quantitatively accurate,
one has to be sufficiently away from the deconfinement transition.

If there are limitations, there are also strengths of this approach. The effective theory
is three-dimensional and purely bosonic, so simulations are cheap and yield rather
precise continuum limits, leaving little ambiguity in the interpretation of
the results. The approach is good for SU(2) as well as SU(3), and permits
to study any number of quarks with any small or zero mass. Finally, the domain of
parameter space where the approach is applicable contains the phenomenologically relevant
region in which heavy
ion collisions are operating. At and above SPS energies, densities in heavy ion collisions
are estimated to be $\mu_B/T\lsim 4.0$ \cite{exp}, i.e.~a quark chemical potential
$\mu/T\lsim 1.3$, which is well within the range where simulations are feasible. 

\subsection{Equilibrium correlation lengths}

Let us first discuss the physics question that will be addressed in the following.
Consider gauge-invariant, local operators $A(x)$. Static equilibrium physics is
described by euclidean
time averages
\be \label{stateq}
\bar{A}({\bf x})=T \int_0^{1/T} d\tau \,A(\bf{x},-\ii\tau)\,.
\ee
The spatial correlation function of such operators,
\be
C(|{\bf x}|)=\langle \bar{A}({\bf x})\bar{A}(0)\rangle_c\sim \ex^{-M|\bfx|},
\ee
falls off exponentially with distance. The
``screening masses'' $M$ have a precise non-perturbative definition: they are the eigenvalues
of the spacewise transfer matrix in the corresponding
lattice field theory. Physically, they correspond to
the inverse length scale over which the equilibrated medium is sensitive
to the insertion of a static source carrying the quantum numbers of $A$. Beyond
$1/M$, the source is screened and the plasma appears undisturbed.
With these definitions, all screening lengths corresponding to gauge-invariant sources
can be computed on the lattice in principle.

\subsection{Dimensional reduction}

If we are interested in length scales larger than the inverse temperature, $|\bfx|\sim 1/gT\gg
1/T$, the situation simplifies considerably. In this case the integration range
for euclidean time averaging in Eq.~(\ref{stateq}) becomes very small, and the problem
effectively three-dimensional (3d). 

The calculation of the
correlation function $C(|\bf{x}|)$ can be factorized:
the time averaging may be performed perturbatively
by expanding in powers of the ratio of scales $gT/T\sim g$,
which amounts to integrating out all modes with
momenta $\sim T$ and larger, i.e.~all non-zero Matsubara modes, in particular the fermions.
This procedure is known as dimensional reduction \cite{dr}. It is in the spirit of a
Wilsonian renormalization group approach, where an effective action for coarser scales
is derived by averaging over the smaller scales. However, here this step is done perturbatively
and entails two approximations: the first is that the computation is obviously limited to
a finite order in $g$. The second is to neglect higher-dimensional operators, which are
suppressed by powers of the scale ratio\footnote{A method to construct an effective theory by a 
(perturbative) block spin transformation has been proposed in \cite{mack}.}.

The remaining correlation function of 3d fields is then to be evaluated
with a 3d purely bosonic effective action, which describes the physics of the
modes $\sim gT$ and softer.
The screening masses $M$ now correspond to the eigenvalues of the 3d transfer matrix.
Simulation of this reduced problem is easy. Without fermions and one dimension less,
much larger volumes and finer lattices can be considered.
Moreover, 3d gauge theories are superrenormalizable and the coupling scales linearly with 
lattice spacing.
Hence, very accurate continuum limits can be obtained.

The effective theory emerging from hot QCD by dimensional reduction
is the SU(3) adjoint Higgs model \cite{dr,rold,ad} with the action
\bea \label{actc}
        S &= &\int d^{3}x \left\{ \frac{1}{2} \tr(F_{ij}F_{ij})
        +\tr(D_{i}A_0)^2 \right. \nn\\
& & \hspace*{1cm}\left. +m_3^2 \tr(A_0^2)
        +\lambda_3(\tr(A_0^2)^{2} \right\} .
\eea
As 4d euclidean time has been integrated over,
$A_0$ now appears as a scalar in the adjoint representation.
The associated Hamiltonian respects
SO(2) planar rotations, two-dimensional parity $P$,
charge conjugation $C$ and $A_0$-reflections $R$.

The parameters of the effective theory are
via dimensional reduction functions of the 4d gauge
coupling $g^2$, the number of colours $N$ and flavours $N_f$, the fermion masses
and the temperature $T$.
In all of the following fermion masses are assumed to be zero,
but in principle any other values may be considered as well.
At leading order in the reduction step one has
\bea 
g_3^2&=&g^2(\bar{\mu})T,\nn\\
m_3&=&\left(\frac{N}{3}+\frac{N_f}{6}\right)^{1/2} g(\bar{\mu})T,\nn\\
\lambda_3&=&\frac{1}{24\pi^2}(6+N-N_f)g^4(\bar{\mu})T\,.
\label{params}
\eea
Note that $m_3$ is just the leading order Debye mass \cite{bel}. 
The reduction step has been performed up to two-loop order \cite{ad} at which parameters
have relative accuracy ${\cal O}(g^4)$
and 1-loop scale dependence on $\bar{\mu}$ is cancelled. 
Specifying $T/\lambdamsbar, N_f$ then completely fixes the effective theory.

If we are interested in 
the infrared-sensitive, non-perturbative scales of order
$\sim 1/g^2T$ \cite{linde},
the fields $A_0 \sim gT$ may be integrated out as well, leaving us
with a 3d pure gauge theory. In order to decide which is the correct effective theory for 
a given temperature beyond parametric considerations,
simulation results for screening lengths will be compared
between the reduced and the full theory.

With the simulation part being rather accurate,
the main error in the correlation functions is due to the reduction step,
and at two-loop level this is formally of the order \cite{rules}
\be
\frac{\delta C}{C}\sim {\cal{O}}(g^3)\,.
\ee
In the treatment of the electroweak phase transition, this is much less than 5\%
\cite{ew}, but for hot QCD the coupling and thus the error is larger. 
In fact, one may wonder why perturbative reduction should work at all, 
given that the effective $g^2\sim 2.7$ at $T=2T_c$ \cite{ad}.
In the loop expansion, the coupling appears as $\alpha_s/\pi$ rather than $g$. 
Inspecting the computed series for 
the effective theory parameters Eq.~(\ref{params}) up to next-to-leading order,  
it appears that it
converges surprisingly well at temperatures not much above the critical one \cite{ad,conv}.
The other source of error is neglecting higher dimension operators. This 
is justified as long as the dynamical mass scales of the 
effective theory are smaller than those integrated out $\sim 2\pi T$. The success of the approach
then depends on whether there is a non-perturbative scale separation.
This can only be checked after simulating the effective theory.

\subsection{Numerical results}

\begin{figure}[tb]%

\begin{center}
\includegraphics[width=6cm]{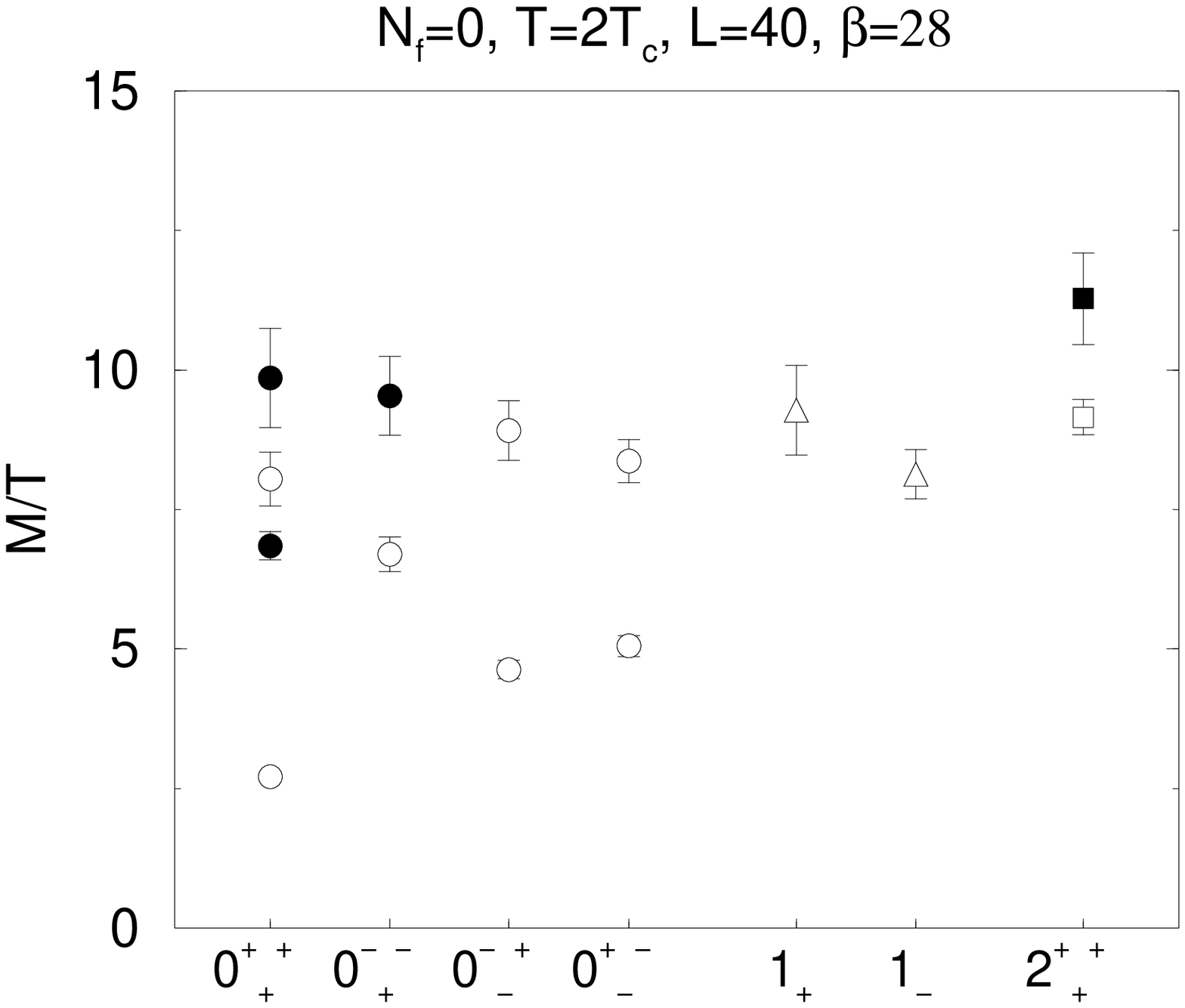}
\end{center}
\begin{center}
\includegraphics[width=6cm]{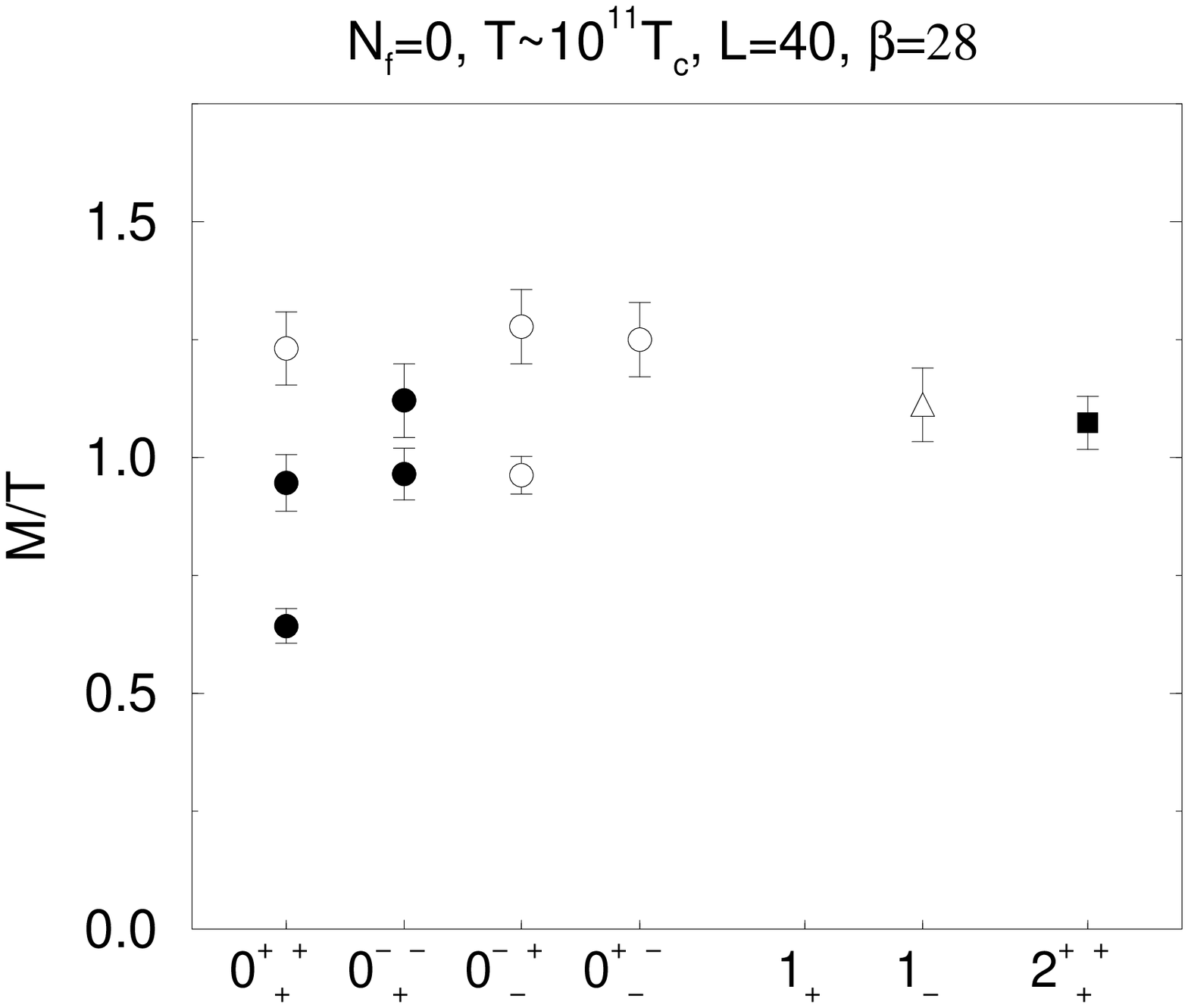}
\end{center}
\vspace*{-1cm}
\caption[]{\label{mt} Spectrum of screening masses in various quantum number channels
at low and high temperatures, for SU(3), $N_f=0$. Filled symbols denote 3d glueball states.
From \cite{hlp1}.}
\end{figure}

Results from simulations \cite{hlp1} of the gauge-invariant screening spectrum with the
effective theory Eq.~(\ref{actc}) are displayed in Fig.~\ref{mt}.
Screening masses are classified according to the symmetries of the 3d Hamiltonian by the
quantum numbers $J^{PC}_R$.
The open symbols show states dominated by scalar operators such as
$\tr (A_0^2), \tr (A_0F_{12}^2)$ etc., whereas full symbols show states receiving
only gluonic contributions $\tr (F_{12}^2)$ etc. It is a remarkable finding of
detailed mixing analyses in several models \cite{mods,hp,hlp1},
that the latter are quantitatively
consistent with the 3d glueball states \cite{teper} and apparently completely
insensitive to the presence of the $A_0$ scalar field. Because of this pronounced
non-mixing, we can thus state a first important result: For any reasonable temperatures
the largest correlation length of gauge-invariant operators belongs to
the $A_0$ degrees of freedom and not to the $A_i$, in contrast to the naive parametric
picture. This demonstrates that the physics from the scale $gT$ down to $g^2T$ is
completely non-perturbative, and hence $A_0$ may not be integrated out perturbatively.
Only for temperatures $T\gsim 10^7T_c$ becomes the coupling small enough that the
parametrically expected ordering of screening masses is realized.

\subsection{Testing dim.~red.~in hot Yang-Mills}

\begin{figure}[bht]%
\vspace*{-0.5cm}
\begin{center}
\includegraphics[width=6cm]{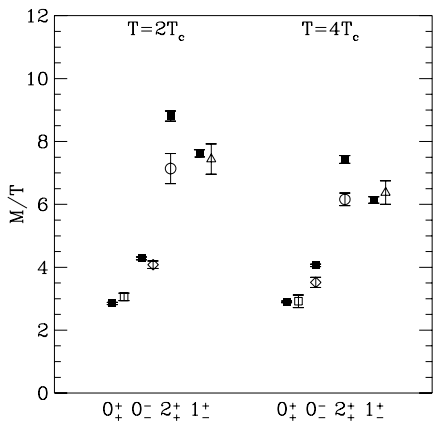}
\end{center}
\vspace*{-1cm}
\caption[]{\label{spec} Comparison of screening masses $J^P_R$ ($C=+$) in hot SU(2) 
pure gauge theory as determined in 4d (empty symbols) and 3d (full symbols)
effective theory simulations. From \cite{hp}.}
\end{figure}
We now wish to know the accuracy of the reduced theory.
Fig.~\ref{spec} compares the results for hot SU(2) gauge theory as obtained
in the 4d lattice theory \cite{dg} with those from the effective theory \cite{hp}. Note that
the effective theory is only valid up to its cut-off $M/T\sim 2\pi$ and above this
level disagreement is to be expected.
The apparent excellent agreement for the two lowest lying states should be taken
with some care given some systematic uncertainties. In 4d simulations,
the projection properties of operators and the scaling behaviour 
are more difficult to control than in 3d. In the effective theory, there is an ambiguity
in the choice of renormalization scale. The direct comparison may thus still suffer 
from effects up to 20\%. 
For the case of SU(3) one finds
again quantitative agreement in the largest correlation length but indeed about 20\%
deviation in the next shorter one, cf.~Table \ref{su3}.
Good agreement between the 4d and the 3d effective theory is also found with other observables,
like the static potential \cite{rold}, the spatial string tension \cite{bali}, the Debye mass \cite{debye}
and gauge-fixed propagators \cite{kar}.\footnote{Dimensional reduction also works to high precision 
from 3d finite $T$ to 2d \cite{bial}.}

Thus, dimensionally reduced pure gauge theory gives at least a semi-quantitative description
of the largest correlation lengths in the system down to temperatures as low
as $\sim 2T_c$.
\begin{table}
\caption[]{Comparison of the lowest two screening masses in SU(3) Yang-Mills at $T=2T_c$,
obtained in 3d \cite{hlp1} and 4d \cite{dg} simulations.}
\label{su3}
\begin{tabular}{|c|c|c|}
\hline
$M/T$      &  3d  & 4d\\
\hline
$0^{++}_+$ &  2.71(6)        & 2.60(4)    \\
\hline
$0^{-+}_-$ &  4.6(2)         & 6.3(2),     \\
\hline
\end{tabular}
\end{table}

\subsection{Inclusion of fermions}

\begin{figure}[tb]%

\begin{center}
\includegraphics[width=6cm]{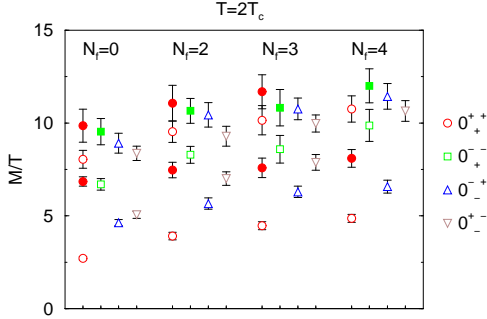}
\end{center}
\vspace*{-1cm}
\caption[]{\label{nf} Spectrum of scalar screening masses 
for various $N_f$ with zero fermion masses. Filled symbols denote 3d glueball states.
From \cite{hlp1}.}
\end{figure}

The main advantage of the effective theory is that fermions, having always non-zero Matsubara 
frequency, are treated analytically. Any change in
the number of fermion species or their masses is encoded in the parameters of the
effective theory, Eq.~(\ref{params}).
The results for the spin zero channels at various $N_f$ (all massless) are shown in
Fig.~\ref{nf}.
The screening masses get larger with increasing $N_f$, so that even
the smallest of them comes close to the cut-off $\sim 2\pi T$ of the effective theory, 
thus diminishing the scale separation needed for the effective theory to be valid. Consequently,
only the lowest state may be expected to 
describe hot 4d physics. As in the $N_f=0$ case, this situation is improved by 
going to larger temperatures.

\begin{figure}[tb]%

\begin{center}
\includegraphics[width=6cm]{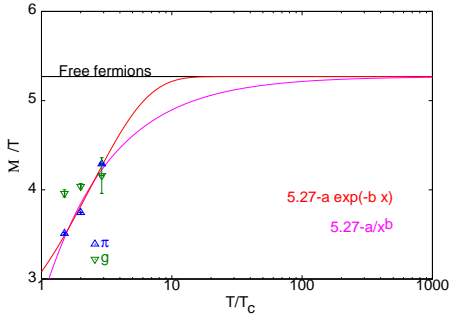}
\end{center}
\vspace*{-1cm}
\caption[]{\label{cross} The smallest scalar screening masses due to correlations of
gluonic (g) and mesonic ($\pi$) operators. The lines represent phenomenological fits.
From \cite{gg}.
}
\end{figure}

Another, and more severe, limitation comes from the fact that, close to $T_c$, the
fermionic modes begin to feel the chiral phase transition and thus become very light through
non-perturbative effects. At some point this effect will be so large that they constitute
the lightest degrees of freedom and may no longer be integrated out. This situation
has been demonstrated recently in a 4d simulation with $N_f=4$ light fermions \cite{gg},
and is shown in Fig.~\ref{cross}.
Below $T\sim 2-3 T_c$, the pion correlator yields the longest correlation length,
and the purely bosonic effective theory fails to give the correct infrared behaviour. This
problem becomes worse for larger $N_f$, whereas it is slightly better for the realistic case
of two light fermion flavours. It would be very interesting to see if this behaviour can
be reproduced in the effective theory, if NRQCD methods are applied and fermionic degrees of
freedom are kept in the effective action \cite{nrqcd}.

\section{DIMENSIONAL REDUCTION AT FINITE DENSITY}

Having seen that dimensionally reduced QCD gives a reasonable description for the longest 
correlation length at temperatures $T\gsim 2-3 T_c$, we may now return to
the problem of introducing a chemical potential \cite{hlp1}. 
The contribution of the fermion determinant with $\mu\neq 0$ 
to the effective potential of $A_0$ has been evaluated
at the two-loop level \cite{kps}. From this one may derive the corrsponding changes in the
three-dimensional effective action, brought about by the chemical potential term. 
To leading order, it generates one extra term in the action Eq.~(\ref{actc}) and changes
the bare $A_0$ mass parameter,
\bea \label{change}
S &\to& S + \ii z\int d^3x\,\tr A_0^3, \quad z=\frac{\mu}{T}\frac{N_f}{3\pi^2},\nn\\
m_3&\to &m_3\left[1+\Bigl( \frac{\mu}{\pi T}\Bigr)^2 \frac{3 N_f}{2 N + N_f}
\right]\,.
\eea
The extra term $S_z$ is odd under $R,C$, and hence the action no longer respects these
symmetries, while parity is left intact. 
Consequently, screening states now are labelled by $J^P$.

\begin{figure}[tb]%
\begin{center}
\includegraphics[width=6cm]{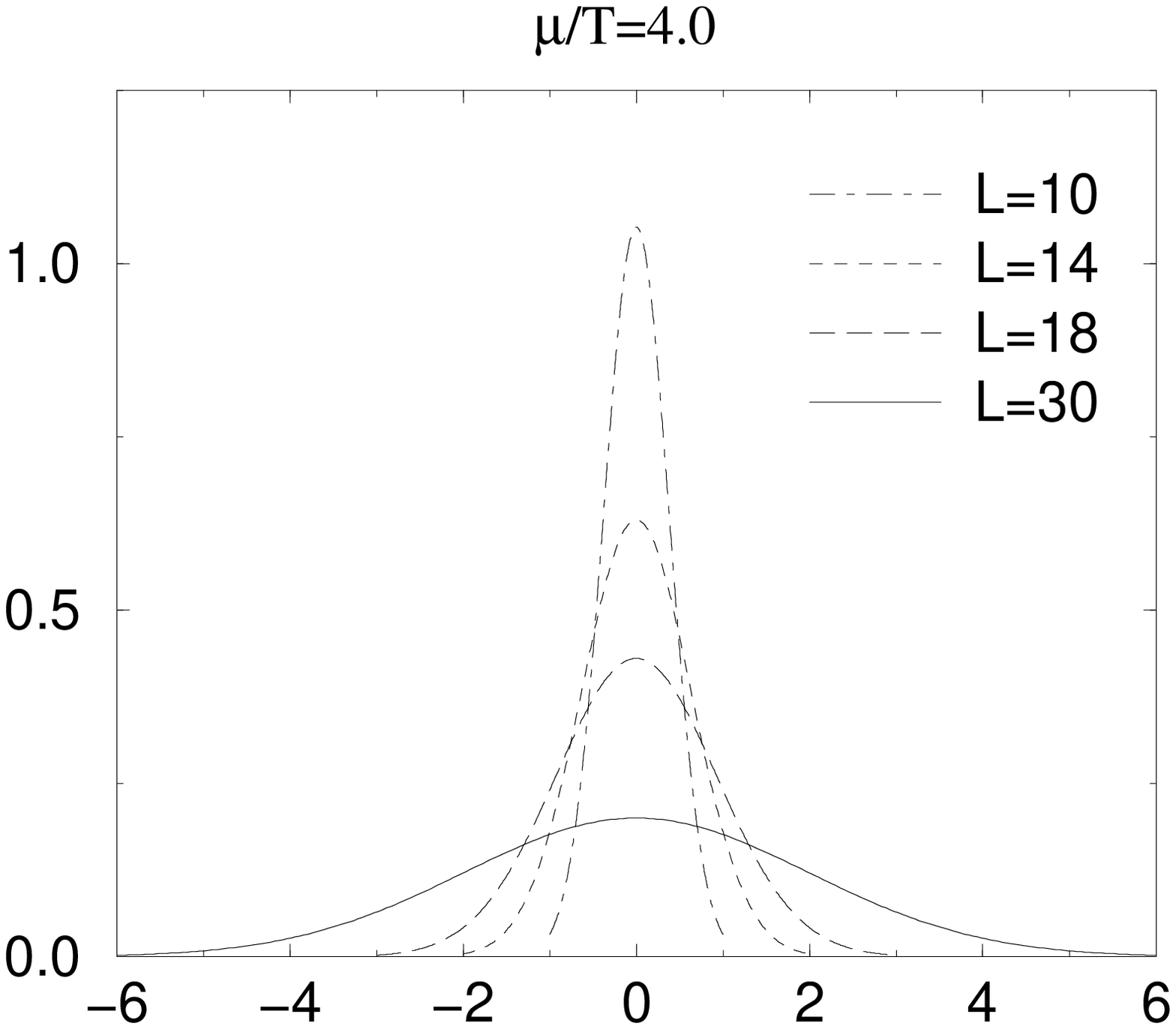}
\end{center}
\begin{center}
\includegraphics[width=6cm]{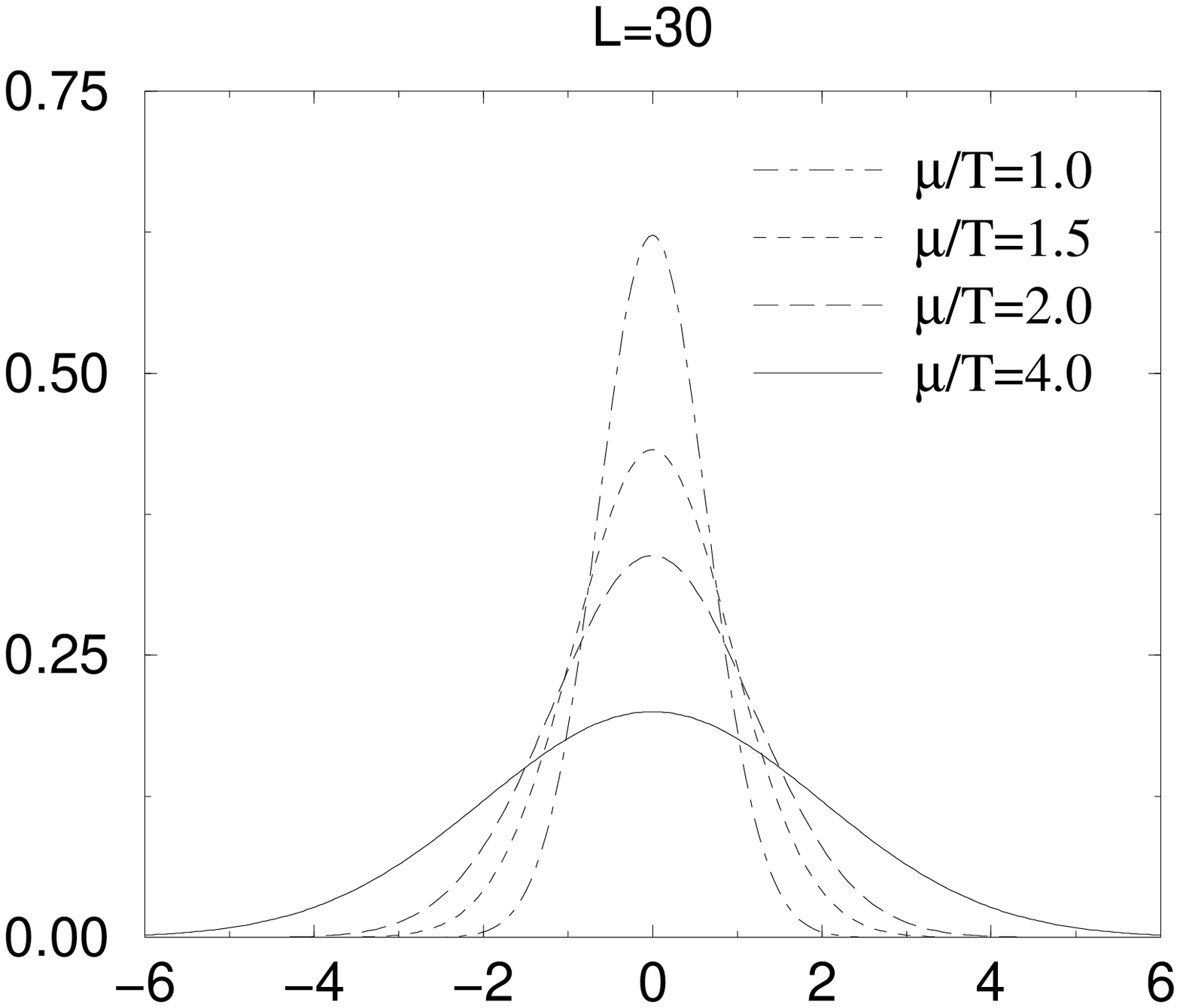}
\end{center}
\vspace*{-0.8cm}
\caption[]{\label{dist} Distribution of the reweighting factor ($zS_z$) for fixed $\mu/T$ and
various $L$ (top), and for fixed $L$ and various $\mu/T$ (bottom). From \cite{hlp1}.
}
\end{figure}

The effective action is complex just as the 4d original one, and hence
there is a sign problem. Expectation values have to be computed
by reweighting with the complex piece of the action
\be
\langle {\cal O}\rangle =
\frac{\langle {\cal O} \ex^{-\ii zS_z}\rangle_0}{\langle\cos(zS_z)\rangle_0}\,,
\ee
where the subscript denotes averaging with $\mu=0$.
Cancelling contributions to the expectation value occur whenever
$(zS_z)\gg 1$. The perturbative width of the distribution at two-loop level can be computed
from the $A_0$ sunset diagram to give
$$\hspace*{-4cm}\Delta (z S_z)
\approx \frac{z}{4\pi}
\Bigl(\frac{6 L}{\beta}\Bigr)^{3/2}$$
\be
\times
\left[5\left(
\ln \frac{1}{12 a m_D}+0.58849
\right)\right]^{1/2}\,.
\label{wi}
\ee
One recognizes the same behaviour as in the 4d case, in that the width of the
phase distribution is growing with $\sim \sqrt{V}(\mu/T)$. 
Fig.~\ref{dist}
shows some distributions of $(zS_z)$ as obtained by Monte Carlo for various volumes and
chemical potentials. 
As long as $\mu\lsim 4T$,
the distribution is well contained within $[-\pi,\pi]$ for volumes large enough
so that the masses attain their infinite volume values  \cite{hlp1}. 
Correspondingly, as long as this condition is met, there are no problems with cancelling
contributions to correlation functions, and the signal is as good as at $\mu=0$. 
For even larger values
of the chemical potential the sign problem sets in and statistical errors explode.

\begin{figure}[tb]%
\begin{center}
\includegraphics[width=6cm]{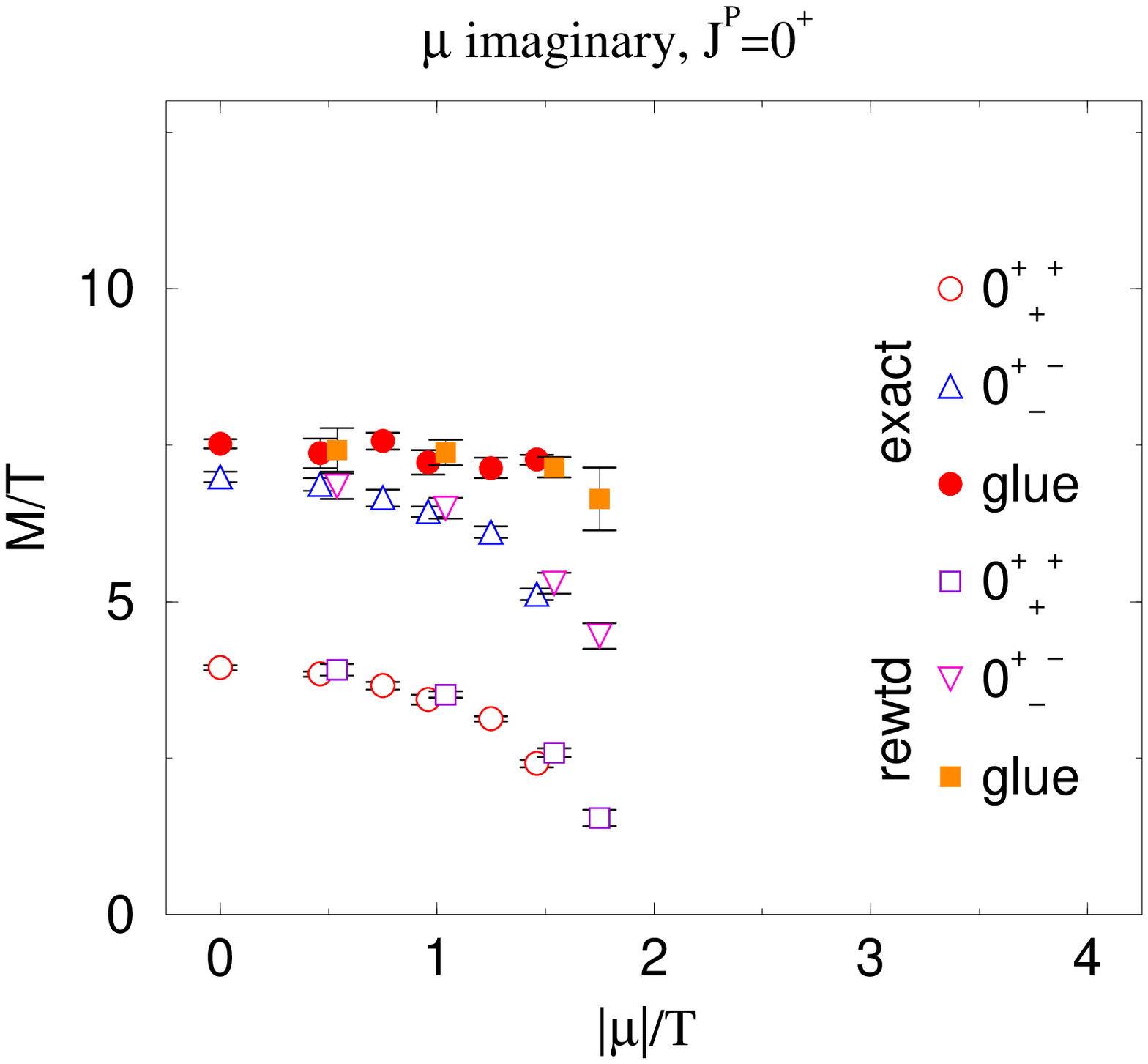}
\end{center}
\vspace*{-0.5cm}
\begin{center}
\includegraphics[width=6cm]{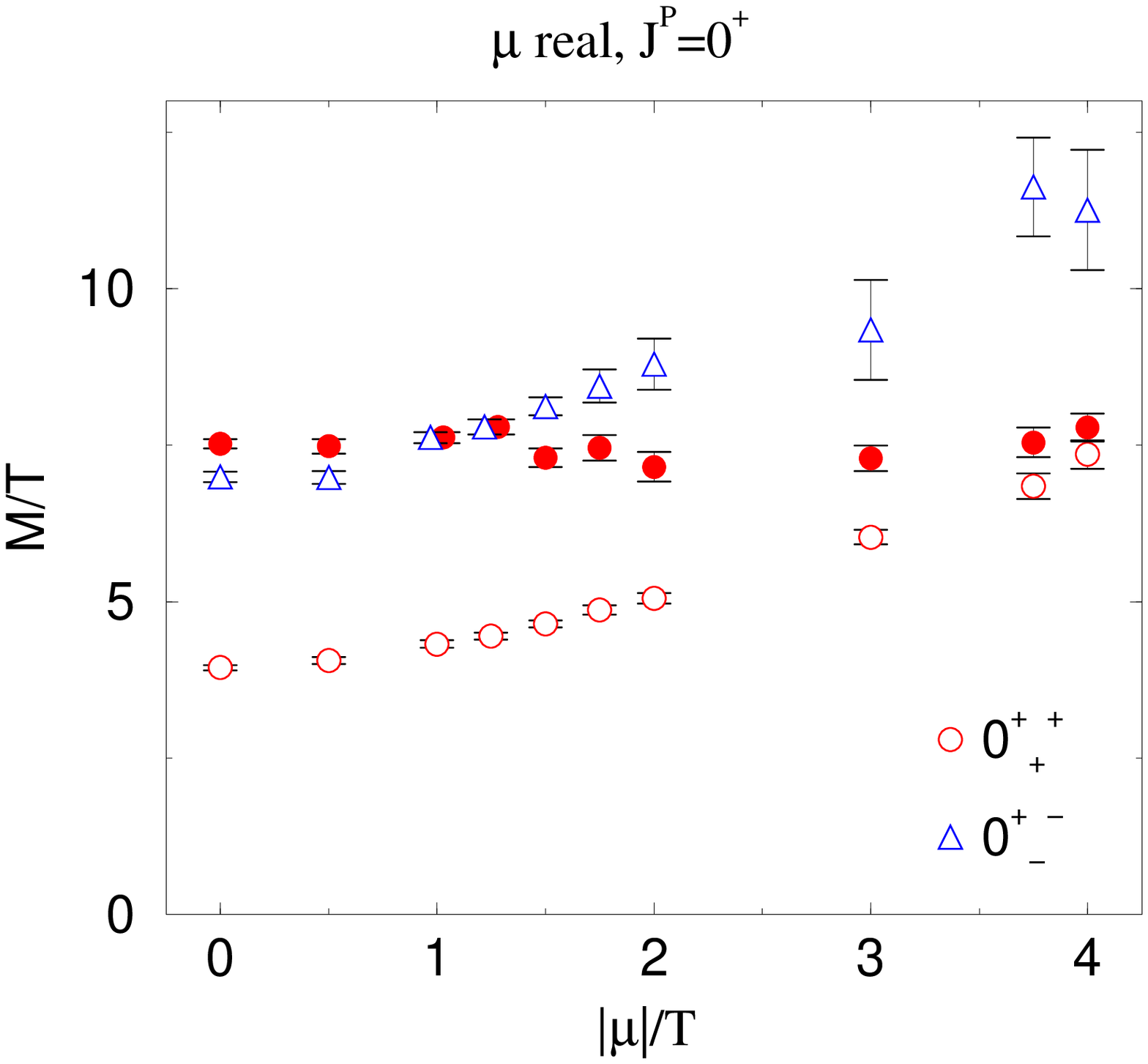}
\end{center}
\vspace*{-1.5cm}
\caption[]{\label{res} The lowest scalar screening masses when an imaginary (top) or real
(bottom) chemical potential is switched on. From \cite{hlp1}.
}
\end{figure}

Next, one has to check the overlap between the
ensembles at $\mu\neq 0$ and $\mu=0$ in order to avoid a wrong biasing of the importance sampling
by the reweighting procedure.
For a given $|\mu|/T$, one can get an idea about this
using imaginary chemical potential, where reweighting can be compared
with the full action update.
This is shown in the top panel of Fig.~\ref{res}. Within the small statistical errors,
no difference between reweighted and original ensembles is noticable.
This result then gives us confidence in the corresponding ones obtained for real
chemical potential, bottom of Fig.~\ref{res}. 

Because of the different qualitative behaviours of 3d gluonic and
scalar states mentioned earlier, we may expect to observe a
change in the nature of the ground state excitation
at some real value of $\mu$.
Indeed, Fig.~\ref{res} bottom suggests a level crossing at $\mu/T\sim 4.0$.
This would
mean that the longest correlation length in the thermal system does
not get arbitrarily short with increasing density, but rather stays at
a constant level. Note that the value of $M/T$ at this crossing is already
so large that the effective theory
may be inaccurate quantitatively.
However, the qualitative effect should be the same in the 4d theory.

\section{RELATION TO 4D}

One may now ask why the sign problem in the effective theory is weakened, and what the relation
to the original problem in four dimensions is. To this end, note that the fermionic Matsubara
modes at finite temperature have propagators $(p_i\gamma_i+m+\gamma_0(2n+1)\pi T)^{-1}$. 
For large $T$ these
modes are heavy, which is why dimensional reduction integrates them out.
We can then make contact to the heavy quark expansion of Sec.~\ref{sec:stat},
where a correlation between the phase of the fermion determinant and the imaginary
part of the Polyakov loop has been observed (cf.~Fig.~\ref{corr}). Expanding the exponential
of the imaginary part of the Polyakov loop  
$P_i \sim {\rm Im}\tr(\int d\tau A_0)^3+\ldots$ indicates that after integrating over euclidean time 
the leading contribution is
\be
\phi \sim \frac{\mu}{T}S_z\;,
\ee
which is just the extra term in the effective action Eq.~(\ref{change}). 
There, part of the path integral
has been computed already analytically, namely the integration over the hard modes
of order $p\geq T$. This has produced the numerical factor in $z$,
Eq.~(\ref{change},\ref{wi}),
suppressing the width of the distribution relative to the one before this integration. 

\section{IMAGINARY VS. REAL CHEMICAL POTENTIAL}

\begin{figure}[tbh]%
\begin{center}
\includegraphics[width=5cm]{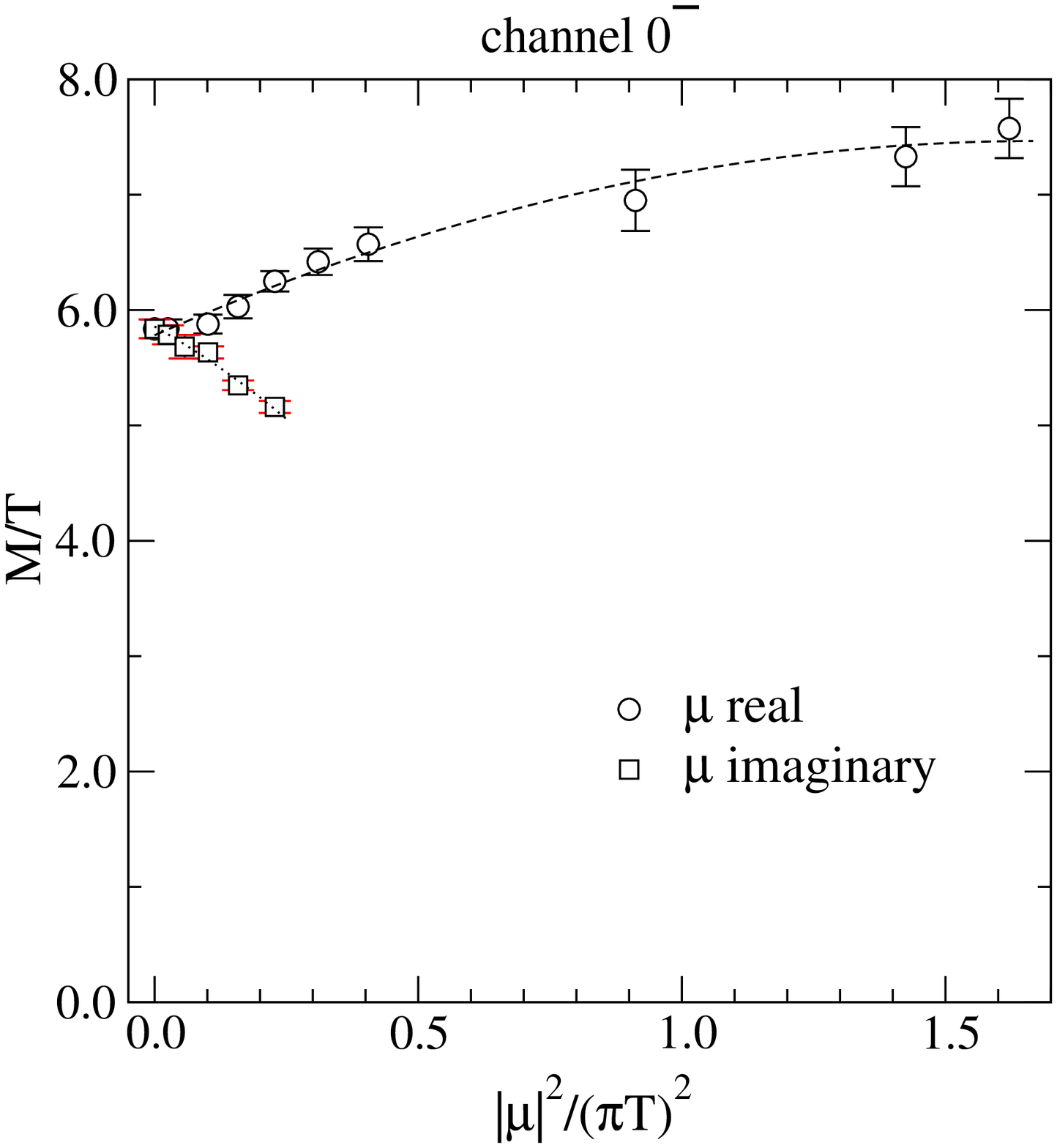}
\end{center}
\begin{center}
\includegraphics[width=5cm]{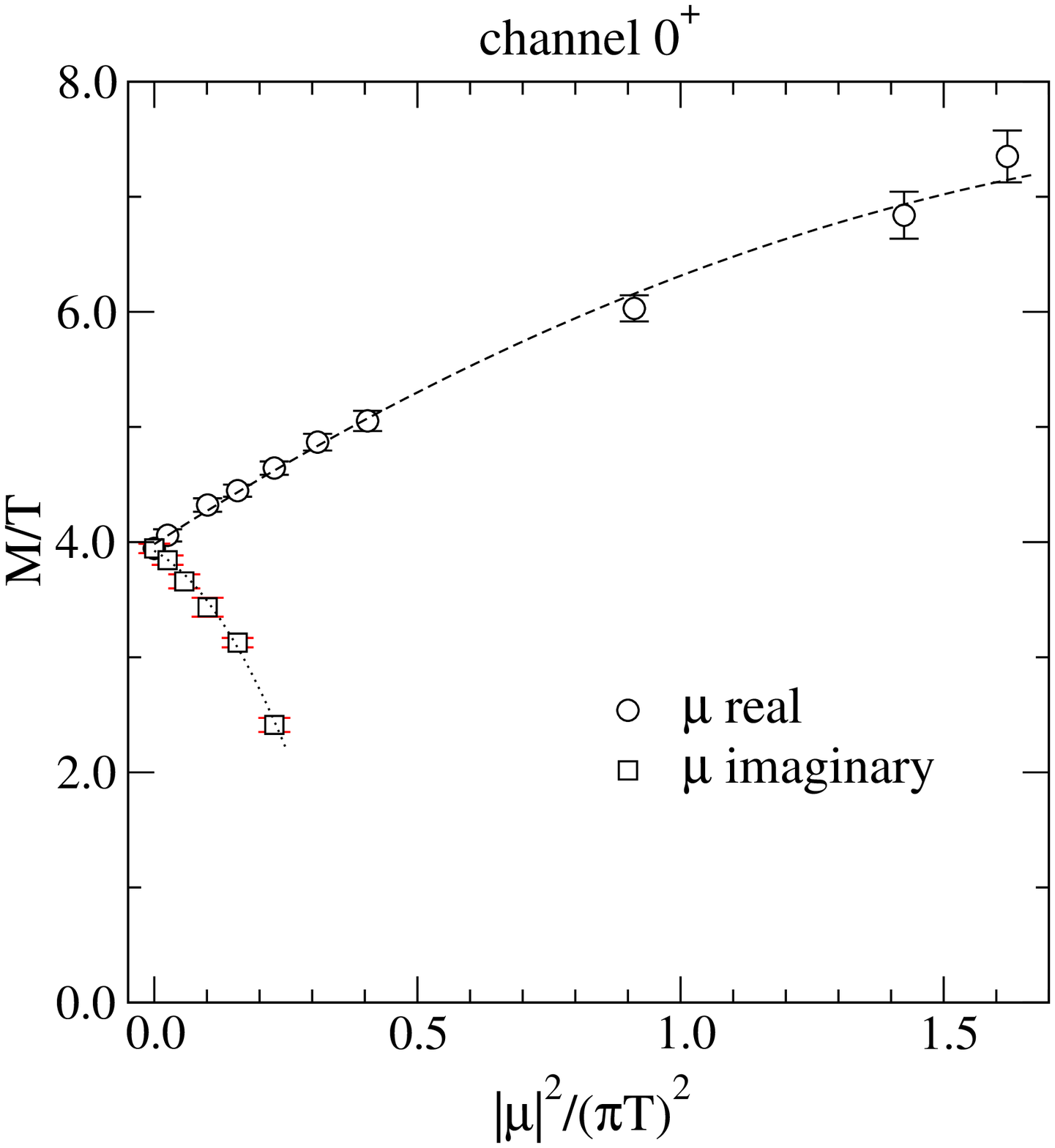}
\end{center}
\vspace*{-1cm}
\caption[]{\label{cont} Fits of Eq.~(\ref{ansatz}) to the lowest two screening masses
at imaginary and real chemical potential. From \cite{hlp2}.
}
\end{figure}
With an effective theory permitting simulation of real and imaginary chemical potential 
at hand, one may now
return to an earlier suggestion and study the feasibility
of analytic continuation of observables from imaginary to real $\mu$ \cite{hlp2}. 
Away from phase transitions, the screening masses
have to be analytic in $\mu/T$, as there are no massless modes in the theory. Moreover,
since a change $\mu\rightarrow -\mu$ can be compensated for by a field
redefinition $A_0 \rightarrow -A_0$ in Eq.~({\ref{actc}), all physical
observables must be even under this operation. In the original 4d
theory the same statement follows from compensating $\mu\to -\mu$ by a
C (or CP) operation. For small values of $\mu/T$,
the screening masses may thus be written
as
\be 
\frac{M}{T}  =  c_0
+ c_1 \biggl( \frac{\mu}{\pi T} \biggr)^2
+ c_2 \biggl( \frac{\mu}{\pi T} \biggr)^4\nn\\
+ {\cal O} \left( \frac{\mu}{\pi T} \right)^6 .
\label{ansatz}
\ee
One may now ask to what extent a truncated power series like this reproduces the data.
In the range where this is possible for both real and imaginary $\mu$, 
the possibility of analytic continuation is easily checked by examining whether
the $c_i$ agree between the two data sets.
Fig.~\ref{cont} shows the lowest states in the system for real and imaginary potential,
together with the fits to Eq.~(\ref{ansatz}).  

Note that in the imaginary $\mu$ case, data are only available up to $|\mu|/T\sim 1.5$,
and hence $c_2$ is not well constrained.
The reason is that imaginary chemical potentials favour a $Z_N$-broken
minimum over the symmetric one in the 4d effective $A_0$ potential, 
once $|\mu|/T\geq\pi/3$.
Thus a phase transition occurs and analyticity is lost.
In the effective theory, this transition shows up by a decrease of the scalar bare mass parameter
sending the system into a broken phase.
In the region up to this critical value, however, complete compatibility for
the values of $c_{0,1}$ is observed between the two data sets \cite{hlp2}.

One may object that, once the deconfinement transition line is continued to complex values
of $\mu$, it may get distorted, and a continuation might fail if it pierces the
transition sheet $T_c(\mu)$ while rotating from imaginary to real values. 
In fact, for imaginary $\mu$ screening masses are decreasing. Correspondingly we would expect the
line $T_c(\mu)$ in the $\{T,{\rm Im}(\mu)\}$-plane to be an increasing function of $|\mu|$, which
is the distortion we were worried about. 
However, the change in $T_c(\mu)$ should be of the same order in $\mu/(\pi T)$ as the 
change in the screening masses. For the maximal $|\mu|/T\leq 1.3$ allowed, the change should
not be more than 20\%, and the procedure is surely safe for $T\gsim 1.5T_c$.

In the region where the effective theory is applicable, it is thus found that direct
analytic continuation does seem to provide a working tool for
determining correlation lengths. This suggests that in 4d simulations
analytic continuation from imaginary chemical potential would
give physical results
if a good ansatz for the $\mu$-dependence is available,
and would allow to go closer to $T_c$ determining, e.g.,
the free energy density and the spatial correlation lengths there.

\section{CONCLUSIONS}

In summary, numerical simulations of SU(3) QCD at finite density are feasible only for static or 
heavy quarks,
where the sign problem persists but enough computational power is available to calculate the phase
of the fermion determinant with sufficient accuracy.
However, all numerical approaches so far fail for realistic QCD.

Qualitative insight may be gained from simulations of two-colour QCD, where the model with
one flavour of adjoint fermions shares several features with QCD,
or from simulations of QCD at finite isopsin density, in which cases there is no sign problem.
It will be interesting to see further developments of another approach,
which attempts to actually solve the sign problem by use of cluster algorithms.  

At temperatures above the deconfinement transition, the equilibrium physics of QCD is described by a 
perturbatively derived, three-dimensional and purely bosonic effective theory, 
which is the more accurate the higher the temperature.
Reasonable results for the longest correlation length may be obtained 
for temperatures $\gsim 2-3 T_c$. 

The effective theory can be efficiently simulated
even after introduction of a finite chemical potential for light or massless quarks, as long as $\mu\lsim 4T$.
This domain of applicability covers the parameter values which are phenomenologically relevant for
heavy ion collision experiments. It is furthermore found that screening masses
can be computed by analytic continuation of calculations at imaginary chemical potential. This points
to the possibility
that static plasma properties could be computed
in four-dimensional QCD simulations employing imaginary chemical potential, thus circumventing the sign problem.
Another problem for future work might be to improve the effective theory by NRQCD methods, allowing also
hadronic observables to be computed.

\end{document}